\documentclass[sigconf,nonacm]{acmart}

\usepackage{graphicx}
\usepackage{booktabs}
\usepackage{multirow}
\usepackage{xcolor}
\usepackage{amsmath} 
\usepackage{enumitem}

\setcopyright{none}
\renewcommand\footnotetextcopyrightpermission[1]{}
\acmConference{}{}{}
\acmYear{}
\copyrightyear{2026}

\begin{document}

\title{GenRec: An LLM‑Backed Recommendation Ranker at Netflix}

%% Single-blind: author names visible, no anonymity required.
%% acmart's native author macros are designed for multiple separate \author blocks.
%% Below we use a single grouped author listing (ACL-style) to match the requested layout.

\author{Ying Li}
\email{yingl@netflix.com}
\affiliation{%
\institution{Netflix}
\city{Los Gatos}
\state{CA}
\country{USA}
}

\author{Shradha Sehgal}
\authornote{Work done while employed at Netflix.}
\email{shradhasehgal7@gmail.com}
\affiliation{%
\institution{Netflix}
\city{Los Gatos}
\state{CA}
\country{USA}
}

\author{Arjun Rao}
\email{arjunr@netflix.com}
\affiliation{%
\institution{Netflix}
\city{Los Gatos}
\state{CA}
\country{USA}
}

\author{Rein Houthooft}
\authornotemark[1]
\email{rein.houthooft@gmail.com}
\affiliation{%
\institution{Amazon}
\city{Sunnyvale}
\state{CA}
\country{USA}
}

\author{Yunan Hu}
\email{yunanh@netflix.com}
\affiliation{%
\institution{Netflix}
\city{Los Gatos}
\state{CA}
\country{USA}
}

\author{Yaochen Zhu}
\email{yzhu@netflix.com}
\affiliation{%
\institution{Netflix}
\city{Los Gatos}
\state{CA}
\country{USA}
}

\author{Sourabh Medapati}
\email{smedapati@netflix.com}
\affiliation{%
\institution{Netflix}
\city{Los Gatos}
\state{CA}
\country{USA}
}

\author{Yun Li}
\email{yli@netflix.com}
\affiliation{%
\institution{Netflix}
\city{Los Gatos}
\state{CA}
\country{USA}
}

\author{Linas Baltrunas}
\email{lbaltrunas@netflix.com}
\affiliation{%
\institution{Netflix}
\city{Los Gatos}
\state{CA}
\country{USA}
}

\author{Grace Huang}
\email{ghuang@netflix.com}
\affiliation{%
\institution{Netflix}
\city{Los Gatos}
\state{CA}
\country{USA}
}

\author{Ashish Rastogi}
\email{arastogi@netflix.com}
\affiliation{%
\institution{Netflix}
\city{Los Gatos}
\state{CA}
\country{USA}
}

\author{Kamelia Aryafar}
\email{karyafar@netflix.com}
\affiliation{%
\institution{Netflix}
\city{Los Gatos}
\state{CA}
\country{USA}
}

\renewcommand{\shortauthors}{Li et al.}

\begin{abstract}
Large language models (LLMs) are reshaping recommender systems by enabling richer modeling of users, content, and context directly in natural language. At Netflix, we are exploring this direction through \textbf{GenRec}, an LLM‑backed recommendation ranker built on top of an in‑house foundational LLM. GenRec follows a \textbf{two‑phase framework}: Phase 1 adapts an open-source LLM to Netflix data, developing deep understanding of the catalog and member behavior while balancing capabilities such as content understanding and instruction following. Phase 2 post‑trains this foundation model with recommendation‑ranking specific data, labels, and reward signals, aiming to align the ranker with business requirements and long‑term member satisfaction. 

This paper focuses on Phase 2 and the transition from a traditional discriminative ranker with thousands of engineered features to an LLM‑backed ranker driven by verbalized user histories and context. We describe our design for input verbalization and context engineering, post‑training data construction, reward integration, model architecture, and a cost‑constrained serving design based on prefill‑only inference approach. We report results from a large‑scale A/B test in comparing GenRec against the current production ranker model, where we show that a GenRec model trained with substantially fewer Phase‑2 labeled training examples and input signals can achieve statistically significant gains in offline and online metrics. We discuss how LLM‑backed recommenders could shift the recommendation paradigm: from feature engineering to context engineering, and from bespoke architectures to shared foundation backbones. We also outline practical lessons for serving such systems under real‑world resource constraints.

\end{abstract}

\maketitle

\section{Introduction}

The recommendation system is central to the experiences of hundreds of millions of members at Netflix. Our current recommender systems in production generate personalized title recommendations based on large number of hand‑crafted features spanning user, item, and interaction signals, along with extensive higher-order feature interactions, in line with prior large-scale recommendation systems ~\cite{matrixfac, dlyoutube, DeepLearningRec, wide&deep}. They also employ customized, complex architectures tailored to different recommendation needs. For example, specialized networks for modeling feature interactions, transformer-based architectures for capturing dynamic member interests, and multi-task learning architectures powering multiple recommendation tasks ~\cite{sasrec, mtl, din, youtubemtl}.

These models have iterated over many years to generate recommendations for a variety of  content types (movies, series, games, live events, podcasts, etc.) across the Netflix product. While this architecture has become a critical part of our member experience, it has become increasingly difficult to support new business needs. For example, onboarding a new content type or recommendation surface often requires substantial work in feature engineering, model architecture design, infrastructure changes, and experimentation. As a result, the system’s complexity makes it difficult to keep pace with the rapid growth of business needs at Netflix.

Recently, large language models (LLMs) are beginning to reshape how recommendation systems are designed, deployed, and leveraged across the industry ~\cite{surveylargelanguagemodelsforrec, llmpromptrec, genrecsurvey}. Their broad world knowledge and expressive language understanding enable richer modeling of users, content, and context than traditional architectures. This opens up new possibilities of the next generation of recommendations that represent user histories and contextual signals directly as natural language to generate better recommendations with deeper understanding of user preferences and item semantics. It also enables quicker adaptations to new use cases, such as steering recommendations via prompts rather than re-designing new specialized models and pipelines.

However, out-of-the box LLMs are not yet suitable as production-level recommenders out of the box: they tend to over-recommend globally popular content, hallucinate out‑of‑catalog titles, ignore nuanced business constraints, and offer limited personalization. To address these issues, we built \textbf{GenRec}, an LLM‑backed recommendation ranker that post-trains an internal foundation LLM on Netflix corpora. Specifically, we focus on a full-catalog ranking use case, and demonstrate that a LLM-based recommendation ranker is a viable and promising path. GenRec represents \textbf{our first step} toward transitioning from a traditional recommendation stack to an LLM‑native system.

\begin{figure*}[t]
  \centering
  \includegraphics[width=0.95\textwidth]{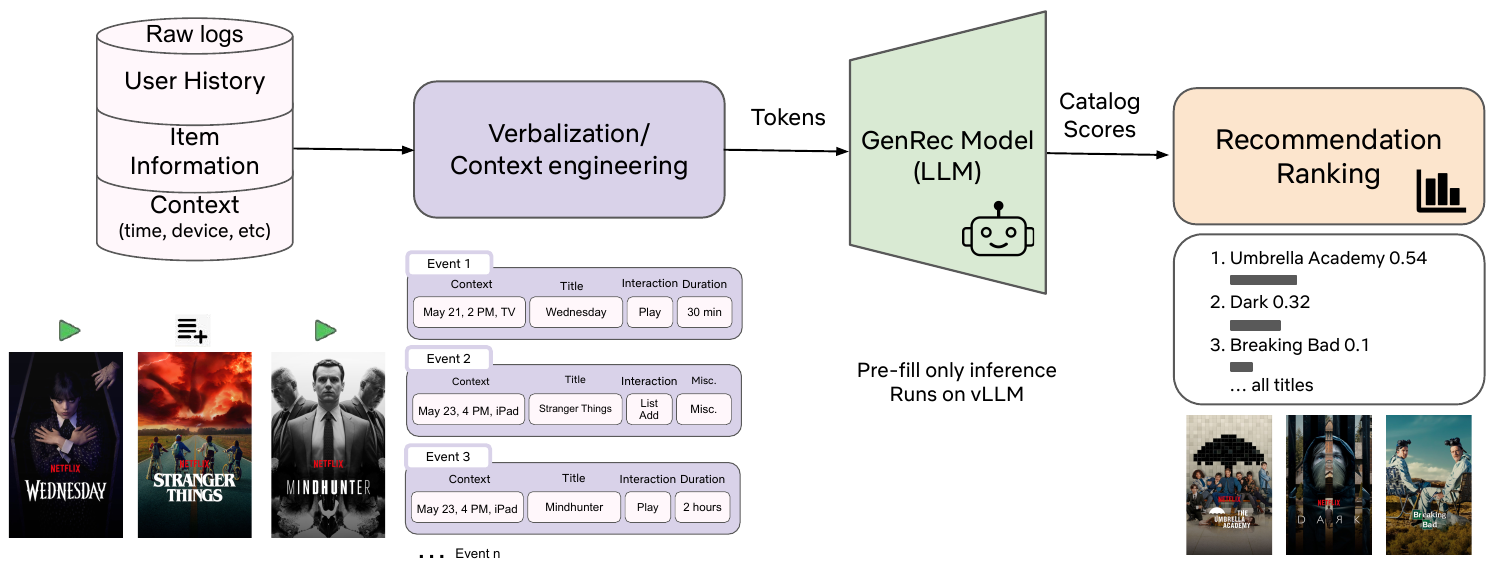} 
  \caption{GenRec inference pipeline. Raw logs of user history, item metadata, and context are transformed via context engineering into natural‑language prompts and fed into the GenRec model, which runs on vLLM in prefill‑only mode and outputs scores for the entire item catalog, inducing a recommendation ranking. \protect\footnotemark}
  \Description{Diagram of the GenRec pipeline showing how user context is verbalized, passed through the LLM in a single prefill pass, and used to rank the full item catalog.}

  \label{fig:pipeline}
\end{figure*}

To generate recommendations with GenRec, we first verbalize user interaction histories, item metadata, and context information as natural language, then tokenize this textual representation and feed it into the LLM (Figure \ref{fig:pipeline}). To learn Netflix‑specific user preferences and avoid over‑recommending popular items, we post‑train the model on Netflix’s catalog and member behavior data. We align the recommendation outputs with long‑term member satisfaction and business objectives by incorporating multiple reward signals during training. We further make the model “Netflix‑catalog‑aware” by constraining its outputs to the items in the Netflix catalog, which eliminates out‑of‑catalog recommendations. Finally, we optimize GenRec under cost constraints and integrate it with Netflix’s LLM serving stack.

  \footnotetext{The verbalization example shown is illustrative only and does not reflect actual user data in production.}

Compared to the existing production recommendation ranker, GenRec substantially simplifies the feature engineering. Instead of manually engineering complex feature interactions, we rely on the LLM model to infer these interactions and user preferences directly from the raw input text. This shifts the recommendation problem from \textbf{feature engineering} to \textbf{context engineering}: deciding what information to present to the model, how to express it within a limited token budget, and what objective and rewards to optimize for, rather than iteratively designing and integrating new features and feature interactions.

GenRec also benefits from the capabilities of the pre‑trained foundation LLM—such as user understanding and content understanding—so we do not need to build these capabilities from scratch. As a result, we observe that GenRec can reach parity with a mature production ranker using substantially fewer Phase‑2 labeled examples and input signals than the traditional models. This marginal data efficiency at the Phase-2 post‑training stage is especially valuable because Phase‑1 foundation training runs at a much lower cadence, whereas Phase‑2 models are refreshed more frequently.

We evaluate GenRec using both offline experiments and a large‑scale online A/B test against a mature production baseline. In the configuration tested, GenRec uses only a small fraction of the Phase‑2 labeled training data and input signals of the current production stack, yet achieves statistically significant improvements on both offline and online metrics, while reducing reliance on hand‑engineered features and bespoke architectures. Together, these results suggest that LLM‑backed rankers are a viable and promising alternative to traditional models, at least on the recommendation surfaces tested. The contributions of this paper include:

\begin{itemize}
 \item \textbf{LLM‑backed ranking that is competitive with a mature production system.} We present GenRec, an LLM‑backed recommendation ranker built on top of an in‑house foundation LLM, and compare it to a long‑running, highly engineered production ranker. A GenRec model trained with substantially fewer Phase-2 training examples and input signals achieves statistically significant gains on our primary offline and online metrics.
 \item \textbf{A two‑phase, LLM‑centric training framework with quantified benefits.} We validate a two‑phase training scheme in which Phase 1 adapts an open‑source LLM to Netflix data to build a Netflix‑aware foundation model, and Phase 2 performs high‑cadence post‑training for recommendation ranking. We articulate the distinct roles and objectives of the two phases and empirically decompose their impact on ranking quality, showing substantial gains from both phases.
  \item \textbf{A LLM‑backed ranking architecture with catalog‑aware scoring.} We describe an architecture that uses a generative, decoder‑only LLM backbone, and a catalog‑aware scoring head that scores large item catalogs in a single forward pass. This design enables large candidate‑set ranking while restricting recommendations to in‑catalog items.
  \item \textbf{Context engineering and reward‑weighted training as practical levers for quality–cost control.} We treat context engineering and reward integration as first‑class design dimensions. On the context side, we show that careful verbalization design can reduce the effective context length to about \textbf{a third} of the original token budget with negligible loss in offline ranking quality, yielding a similar factor reduction in serving cost. On the objective side, we integrate multiple reward signals via a reward‑weighted ranking loss, steering the model toward long‑term value and business goals in a robust and cost-efficient way.
  \item \textbf{Empirical observation on data/model scaling and quality–cost trade‑offs.} We characterize how recommendation quality scales with Phase‑2 post‑training data and with backbone sizes, and how these gains interact with training and serving costs. We observe consistent, monotonic data scaling behavior across model sizes, and identify practical “sweet spots” on the quality–cost Pareto frontier for Phase‑2 configurations. These findings offer actionable guidance for designing LLM‑backed recommenders under resource constraints.
\end{itemize}

The rest of the paper is organized as follows. Section 2 situates our work within existing research on generative and LLM-based recommendation. Section 3 formalizes the problem setting. Section 4 describes our methodology, including input verbalization, model architecture, post-training data and rewards, and serving optimizations. Section 5 presents results from offline experiments and a large-scale online A/B test. Section 6 discusses potential shifts introduced by LLM-native recommendation. Section 7 concludes and outlines directions for future work.

\section{Related Work}

\subsection{Generative Recommendations with Special Tokens}

A growing line of work models recommendation as autoregressive generation over user interaction sequences. Sequential recommenders such as SASRec~\cite{sasrec} and BERT4Rec~\cite{bert4rec} established this template with next-item prediction over sequences of item IDs. Industrial systems have since scaled this paradigm using transformer-decoder backbones whose vocabularies are extended with special tokens encoding items ID, item metadata, locale, time, devices, surfaces, and so on, allowing the model to condition on heterogeneous information~\cite{hstu,netflixfm,pinrec}.

A complementary direction tokenizes items into semantic codes to capture the semantic meaning of items. TIGER~\cite{tiger}, for example, represents each item as a short sequence of ``semantic IDs'' learned via an RQ-VAE over content embeddings, enabling autoregressive decoding over billion-item catalogs. Variants of this approach have been adopted across multiple platforms, including for ranking at Google~\cite{semanticids} and end-to-end retrieval-and-ranking at Kuaishou~\cite{onerec}.

These models have demonstrated strong performance, but they are typically limited in their ability to leverage broad world knowledge, to steer recommendations through flexible prompts, or to inherit the richer capabilities of modern pretrained language models. These limitations have motivated a parallel line of work that builds recommenders directly on top of LLM backbones.

\subsection{LLM-based Recommendation}

The success of large language models in various tasks has inspired a line of work that uses pretrained LLMs as the backbone of recommenders, rather than training recsys-specific transformers from scratch with special tokens. Early academic systems cast recommendation as text-to-text generation on top of language models~\cite{p5,tallrec,llamarec}.

To allow LLMs to generate items directly rather than free-form text, recent industrial systems extend the LLM vocabulary with Semantic-ID item tokens. Google's PLUM~\cite{plum} adapts a pretrained LLM at YouTube scale through Semantic-ID tokenization, continued pre-training on domain-specific data, and task-specific fine-tuning. Spotify's GLIDE~\cite{glide} formulates podcast discovery as instruction-following over a Semantic-ID catalog. Kuaishou's OneRec-Think~\cite{onerecthink} extends OneRec with a Qwen3 model backbone and reasoning steps.

Most LLM‑based generative retrieval systems rely on autoregressive decoding with beam search at inference time, which introduces latency overhead that can be prohibitive at scale, especially for large candidate sets. Our work avoids these limitations by augmenting the base LLM architecture with a catalog-aware ranking head, so that it can rank a large candidate set in a single forward pass, which makes the serving more cost-effective. In addition, we design the system explicitly through an LLM lens, including effective context engineering and reward integration to satisfy business requirements and optimize long-term user satisfaction within real-world resource constraints.

\section{Problem Setting}

In GenRec, we focus on a full‑catalog ranking problem (or top K ranking if candidate set is provided). The resulting ranked list serves as a personalized representation of user preferences that can be reused across diverse use cases, including item ranking on various surfaces and personalized input signals to downstream applications.

\vspace{1mm}

\noindent \textbf{Notation.} Let $U$ denote the set of users, $C$ the catalog of items (movies, shows, games, live events, podcasts, etc.), and $X$ the space of contexts (device, surface, locale, time of day, etc.). Here, we consider a single recommendation request characterized by a user $u \in U$, a context $\tau \in X$, and a time $t$; let $H$ denote the user's interaction history prior to $t$. Each item $i \in C$ is associated with metadata $M_i$ (title name, genre, synopsis, release date, etc.).

\vspace{1mm}

\noindent \textbf{Ranking Target.}
A recommender maps a request ($u$, $\tau$, $t$, $H$) to a ranking $\pi$ of the catalog $C$, $\pi : C \rightarrow \bigl\{ 1, \ldots, |C| \bigr\}$, where $\pi(i)$ denotes the position assigned to item $i$ (with rank 1 at the top of the list). The objective is to select a ranking $\pi$ that maximizes expected long-term member utility --- a proxy for member satisfaction and retention --- rather than short-term engagement alone.

\vspace{1mm}

\section{Methodology}

\subsection{Overview}

\begin{figure}[h]
  \centering
  \includegraphics[width=\linewidth]{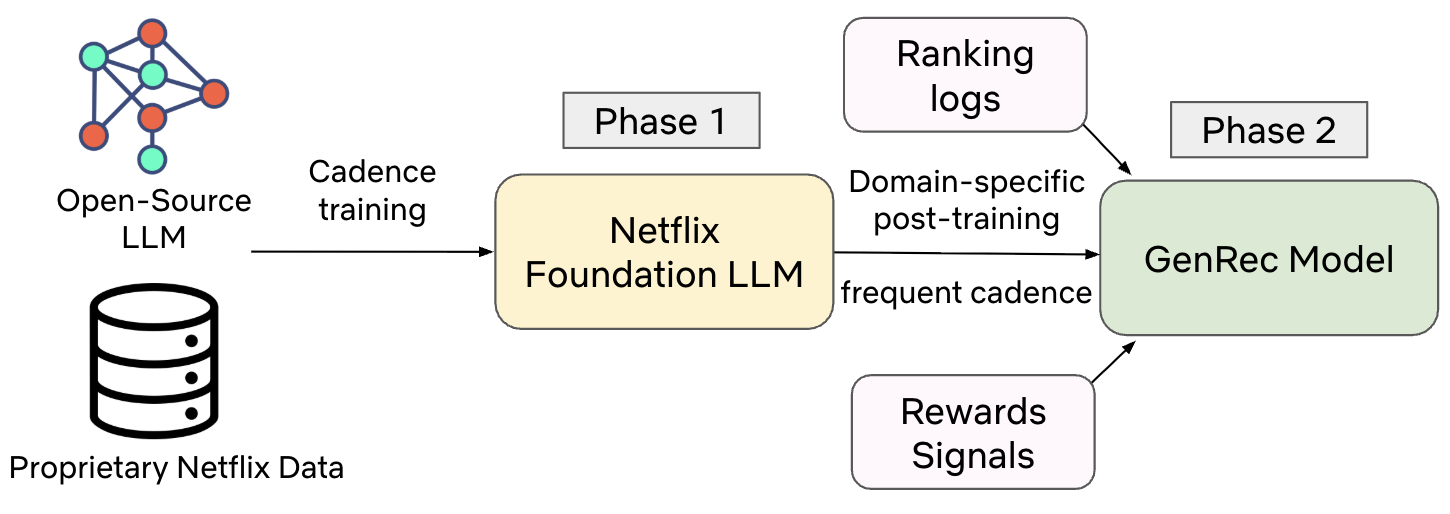}
  \caption{Two-phase training framework. Phase 1 trains a foundational LLM on Netflix data for user and content understanding, and Phase 2 post-trains on recommendation-ranking-specific data and objectives.}
  \Description{Diagram showing the two-phase training framework for GenRec: Phase 1 trains a foundational LLM on Netflix data for user and content understanding. Phase 2 post-trains for recommendation-ranking-specific objectives using ranking data and reward alignment.}
  \label{fig:two-phase}
\end{figure}

GenRec follows a \textbf{two-phase training framework} (Figure ~\ref{fig:two-phase}). In \textbf{Phase 1}, a foundational LLM is trained on top of OSS models with proprietary Netflix data to develop deep understanding of Netflix user and content. In \textbf{Phase 2}, we further post‑train this foundational model on ranking-application specific data and objectives, producing GenRec, which is tailored and optimized for the recommendation ranking task.

Compared to Phase 1, GenRec Phase‑2 post‑training pursues a few different objectives:

\begin{enumerate}
  \item \textbf{Capability focus:} Phase 1 optimizes and balances a broad set of foundational capabilities, such as world knowledge, personalization capability, content understanding, and language capability. Phase 2, by contrast, focuses more on recommendation ranking capabilities, such as ranking quality and recommendation steering. 
  \item \textbf{Update cadence:} Phase 1 is updated relatively infrequently, targeting long‑term user and content understanding and language capabilities. Phase 2 needs to be updated more frequently to track new content launches, evolving popularity patterns, and members’ latest interests.
  \item \textbf{Cost sensitivity:} Phase 1 aims to build the most capable foundational model and is less constrained by serving cost; while Phase 2 must be explicitly cost‑efficient to serve large-scale traffic.
\end{enumerate}

\subsection{Post-Training Data}

At Netflix, members generate \textbf{hundreds of billions} of interaction events across diverse content types, including movies, shows, games, lives, and more. These interactions span a wide range of engagement signals — viewing, playing, thumbing, adding to list, etc. — and occur across multiple recommendation surfaces.  

To post-train LLMs with recommendation capabilities, we transform a member’s historical interaction data into a single‑turn or multi‑turn “conversation” between a user and a recommender. Each turn pairs a \textbf{user message} (combining context, history, and a task) with an \textbf{assistant message} representing the member’s actual engagement for that turn. Specifically, each conversation contains: 

\begin{itemize}
  \item \textbf{Context}: surface, time, device, locale, etc.
  \item \textbf{User profile \& history}: country, tenure, plan, historical interactions, etc.
  \item \textbf{Item-level detail}: item IDs, item metadata (title name, release time, synopsis, etc.), popularity trends, etc.
  \item \textbf{Task}: for example, predict the next items that the user will play or thumb up
  \item \textbf{Assistant message}: the user’s implicit or explicit feedback (play, play duration, abandon, thumb, etc.), which acts as the ground-truth signal during training. 
\end{itemize}

During Phase‑2 post‑training, the LLM is optimized on this conversational data to model how assistant messages depend on the preceding user messages. This conversational framing provides a unified way to express rich recommendation logs as text, which is then used to train both the language modeling objective and the catalog‑aware ranking objective described in the Model Architecture section.

%During \textbf{inference time}, we reuse the same verbalized context (context, profile, history, and item metadata) as input to the LLM to obtain a pooled representation, and then apply the catalog‑aware scoring head to score items. We do not decode full assistant messages at serving time; the conversational format is used primarily during post‑training to support the language modeling objective and maintain strong language understanding of the verbalized text. 

\subsection{Input Verbalization and Context Engineering}

Unlike traditional recommenders that encode user interactions and item metadata into hand‑crafted features or dense embeddings, GenRec \textbf{verbalizes} rich user interaction histories and context as natural language or lightly structured text. By encoding raw interaction signals directly in the LLM’s semantic space, GenRec relies on the LLM to learn higher‑level patterns—such as item relationships, temporal dynamics, and evolving user interests—rather than encoding these relationships explicitly through manual feature engineering.

Verbalization is not merely converting signals into text; it requires deliberate \textbf{context engineering }~\cite{anthropicblog} to accommodate long user context under the finite token budget. A member’s interaction history can easily exceed the token budget, both because of the large number of interactions and the richness of per‑interaction metadata. The context window is a limited and valuable resource, where excessively long contexts can dilute attention and significantly increase training and inference costs.

We therefore carefully curate which events and attributes enter the context window:

\begin{itemize}
  \item \textbf{Retain in full}: high‑signal engagements, such as long‑duration plays and thumbs‑up events, are verbalized with richer metadata. % (e.g., title name, engagement timestamp and device, engagement duration, engagement type, etc.).
  \item \textbf{Omit entirely}: recent, low‑signal events—such as very short plays or noisy views/clicks—are omitted, as they contribute little to ranking quality relative to their token cost.
  \item \textbf{Summarize or compress}: repetitive behaviors, such as binge‑watching sessions, are compressed instead of listing every event repeatedly.  
  \item \textbf{Elaborate selectively}: for important items (e.g. new releases or cold-start items), we can include more detailed metadata to compensate for the limited information available from the pretrained model and historical interactions. 
\end{itemize}

Within the fixed token budget, we prioritize short‑ to medium‑term history with higher granularity, while older history is either omitted or compressed into a brief user‑interest summary. The overall goal is to construct a compact set of highly informative tokens that preserves recommendation quality without exceeding the context window or incurring prohibitive inference costs.

\subsection{Post-Training Objective}

GenRec Phase‑2 training combines mainly two main objectives:
\begin{enumerate}
  \item \textbf{Recommendation ranking objective.} The primary task is a catalog‑aware ranking objective that trains the model to assign higher scores to items that correspond to high‑value engagements. In this task, positive labels correspond to high‑quality engaged events (e.g., high-duration plays,  strong explicit feedback), with denoising logic and thresholds that vary by content types . These labels are used in a cross entropy loss over the item catalog (or candidate set), teaching the model to produce a high‑quality ordering of items given the verbalized context. 
  \item \textbf{Language modeling objective.} In addition, we include a language modeling objective over the verbalized inputs and outputs (such as predicted titles and other textual fields). This task helps maintain the backbone’s general language understanding and text generation capabilities, which are important for (i) modeling rich, natural‑language user histories and item metadata, and (ii) enabling recommendation steering via prompts.
\end{enumerate}

During \textbf{training time}, the model is optimized jointly to (i) score items for ranking and (ii) model the textual domain (e.g., predicting tokens in titles and verbalized context). During \textbf{inference time}, we currently use only the ranking task to generate recommendation ranking. The language modeling objective primarily serves to introduce complementary textual information to improve the ranking task, preserve language understanding and responsiveness to prompt‑based steering, and keep open the possibility of future text generation use cases (e.g., natural‑language explanations).

\vspace{1mm}

The overall GenRec model is trained with a multi-objective loss that combines the recommendation ranking objective, the language modeling objectives, and other objectives where applicable.

 $$\mathcal{L} = \alpha \cdot \mathcal{L}_{\text{ranking}} + \beta \cdot \mathcal{L}_{\text{language}} + \gamma \cdot
  \mathcal{L}_{\text{miscellaneous}}$$
  $$\text{s.t.} \quad \alpha + \beta + \gamma = 1, \quad \alpha, \beta, \gamma \geq 0$$

The $\alpha$, $\beta$, $\gamma$ are tunable hyperparameters which can be tuned through offline experiments. 

\subsection{Model Architecture}

GenRec’s backbone architecture closely follows the foundational LLM: a \textbf{decoder‑only Transformer} trained with next‑token‑prediction style objectives. It is augmented with a \textbf{catalog‑aware ranking head} that constrains recommendations to the Netflix in-catalog items and supports efficient scoring over large candidate sets. 
%This architecture allows GenRec to inherit  the foundational LLM’s knowledge while ensuring that recommendations are restricted to in‑catalog items via the catalog‑aware head.

\vspace{1mm}

\noindent \textbf{Scoring Model.}
The scores are computed in the following three stages: 
\begin{enumerate}[label=\textbf{(\arabic*)}, leftmargin=2.0em, itemsep=0.6em, topsep=0.6em, parsep=0pt]
\item \textbf{Verbalization.} A verbalizer $V$ maps the interaction history, the context, and the item metadata into a single text sequence:
\begin{equation}
              x = V\bigl( H,\ \{M_i\}_{i \in C},\ \tau \bigr)
\end{equation}

\item \textbf{Pooled representation.} The LLM encodes $x$ into a $d$-dimension representation $h$, taken as the hidden state at a pooling position, which summarizes the user preference and context.

\item \textbf{Catalog-aware scoring.} A scoring head $\phi$ assigns each item $i$ a score by combining $h$ with a learned $d$-dimensional embedding $e_i$ as the recommendation ranking task.
\end{enumerate}

\noindent The parameters $\theta$, including the weights of LLM, the scoring head $\phi$, and the item embeddings $\{e_i\}$, are trained jointly. A softmax over the catalog converts these scores into a probability distribution. Afterwards, the catalog ranking $\pi$ can be derived from the item scores. When the catalog is too large to score exhaustively, we can combine the scoring with sampled-softmax instead.

% The backbone and ranking head are trained jointly end‑to‑end in GenRec. Concretely:
% \begin{enumerate}
%   \item The verbalized user-and-context text is fed through the LLM, and we extract a pooled user-context representation $h \in \mathbb{R}^d$ from a designated pooling position in the final layer.
%   \item Each catalog item $i$ has a learned item embedding $e_i \in \mathbb{R}^d$, trained jointly with the backbone.
%   \item A lightweight scoring module $\phi$ combines $h$ and $e_i$ to produce a scalar score $s_i$. In practice, $\phi$ is implemented as a simple function over $h$ and $e_i$ (e.g., dot product and/or a small MLP), chosen to keep scoring efficient over large catalogs.
%   \item Applying $\phi$ across the catalog (or a pre-retrieved candidate set) yields scores $s_i$, which define a distribution over items and induce the final ranking.
% \end{enumerate}

\subsection{Reward Signals}

Beyond raw recommendation accuracy, GenRec must also satisfy two additional goals:

\begin{enumerate}
  \item \textbf{Respect business requirements}, such as content mixing logic (e.g., balance movies, shows, live, games, podcast, and other content types).
  \item \textbf{Maximize long‑term member satisfaction}, rather than just optimizing short‑term user engagements.
\end{enumerate}

Simply post‑training on raw interaction sequences risks drifting away from these goals. For example, the model might over‑recommend binge‑watching over discovery, favor videos over games, or select items that maximize immediate clicks at the expense of catalog exploration or long‑term retention. To address this, we incorporate reward signals derived from separate reward models into GenRec’s training objective. We use two broad categories of rewards:
\begin{enumerate}
  \item \textbf{Long‑term satisfaction proxies.} These rewards estimate how strongly a short-term engagement event correlates with long‑term outcomes. Because true long‑term metrics are noisy and delayed, we rely on more stable proxies learned from historical data. For example, the reward models assign higher values to engagements that encourage members to return to the service, explore across a wider portion of the catalog, or lead to sustained engagement over time.
  \item \textbf{Behavior rebalancing across content and engagement types.} Additional rewards are used to rebalance behavior across content types (e.g., movies vs. games vs. live vs. podcast) and launch stage (e.g. pre-launch, newly launched, ever-green), in order to meet business objectives such as exposure fairness across content types.
\end{enumerate}

We leverage multiple reward models from an existing reward modeling framework ~\cite{rewardinnovation} and integrate their outputs into GenRec via a \textbf{reward-weighted ranking loss}. Concretely, each training example is assigned a scalar weight derived from the multiple reward signals, and the ranking loss for that example is scaled accordingly. High‑value engagements (according to the reward models) receive larger weights, while low‑value or undesirable behaviors are down‑weighted.

This reward‑weighted loss provides a simple, robust, and cost‑effective way to steer the model toward long‑term value and business‑aligned behavior. Compared to full reinforcement learning approaches, this approach is easier to deploy and maintain. In preliminary experiments, RL‑style methods (e.g. GRPO) showed additional gains over supervised finetuning, but their high training overhead makes them better suited for future work. Therefore, currently we adopt the reward‑weighted loss approach as the main alignment mechanism for GenRec, due to its simplicity, stability, and cost considerations.

\subsection{Online Serving and Cost Optimizations}

GenRec is served with Netflix’s internal LLM serving stack using the vLLM framework \cite{netflixserving}. At Netflix scale (hundreds of millions of members globally), serving cost is a first‑class constraint, and inference cost in practice is roughly proportional to the product of model size and context length. We therefore pursue two complementary strategies to optimize the serving cost. First, we \textbf{explore smaller and distilled language models} ~\cite{mobilellmr1} trained on larger or more targeted datasets, aiming to recover much of the quality of larger models while reducing per‑request computation. Second, we aggressively \textbf{optimize context length via verbalization compaction}, as described in the previous section, reducing the tokens used to represent input signals without significantly degrading offline ranking quality.

A further lever on inference cost is the \textbf{inference mode}. Since GenRec needs to rank items over a large candidate set, naïve autoregressive decoding (e.g. generating one token at a time or using beam search) would be prohibitively expensive. Although the underlying backbone is a generative, decoder‑only LLM capable of autoregressive decoding, we intentionally deploy GenRec in a \textbf{prefill‑only }configuration: the model consumes the input context once, and produces ranks for the full candidate set in a single forward pass. This avoids step‑by‑step decoding and makes it feasible to serve GenRec on high‑volume workloads within our compute budget.

\section{Experiments \& Results}

In this section we evaluate GenRec against the production model, analyze how performance scales with data and model size, quantify the contributions of Phase‑1 and Phase‑2 training, and study the impact of context length optimization.

\begin{figure}[h!]
  \centering  \includegraphics[width=\linewidth]{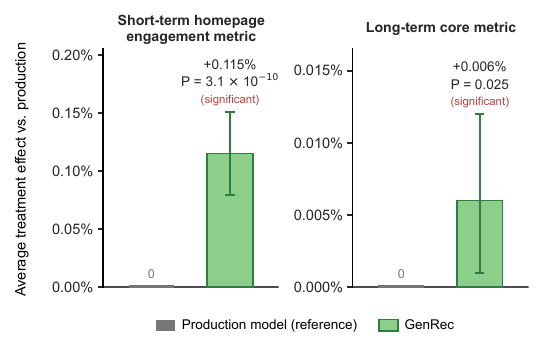}
  \vspace{-2mm}
  \caption{Online metrics of GenRec vs. production model. GenRec achieves statistically significant improvements on both short‑term and long‑term online metrics. \protect\footnotemark}
  \label{fig:online}
\end{figure}
% \footnotetext{The difference in improvement percentage is purely due to the definition of the metric, which does not mean the improvement for short-term engagement is more than the long-term engagement.}

\subsection{Comparison of GenRec vs. Production Baseline}

We compared GenRec with the production baseline, a mature discriminative model tuned over many years. %The baseline model uses thousands of engineered dense and embedding features, with special networks to learn feature interactions, and transformers to model the user sequence. 

Offline, GenRec exceeds the production model on ranking metrics, Mean Reciprocal Rank (MRR), while using significantly less Phase‑2 training data and fewer input signals. Concretely, with about 40× less Phase‑2 labeled training examples than the production model, GenRec achieves an offline lift of approximately +1.6\% relative in MRR over the baseline. Moreover, offline metrics continue to improve as we further scale up the training data and input signals.

We validate this result in a large-scale online A/B test, focusing on key batch‑compute surfaces. We allocated approximately \textbf{~10\% }of Netflix traffic for \textbf{~4 weeks}. Under a low‑data, low-signals configuration, the Phase-2 GenRec model achieves statistically significant improvement over the production baseline on both short-term and long-term online metrics (see Figure \ref{fig:online}). For example, we observe a +0.006\% relative improvement on our core online metric, which is statistically meaningful at Netflix scale. These findings demonstrate that an LLM‑backed ranker, when combined with appropriate post‑training and reward signals, can become a viable and promising alternative to traditional recommendation ranker. %They also suggest substantial headroom as we further scale both data and input signals.

The relatively small number of training samples and input signals used by GenRec, compared to the production ranker, illustrates the marginal data efficiency of Phase‑2 post‑training. Since Phase 2 is updated much more frequently than Phase 1, improving data efficiency at this stage is especially valuable in practice, as it directly reduces the compute required to keep the ranker fresh.

\subsection{Data and Model Scaling}

\begin{figure}[h]
  \centering
  \includegraphics[width=0.9\linewidth]{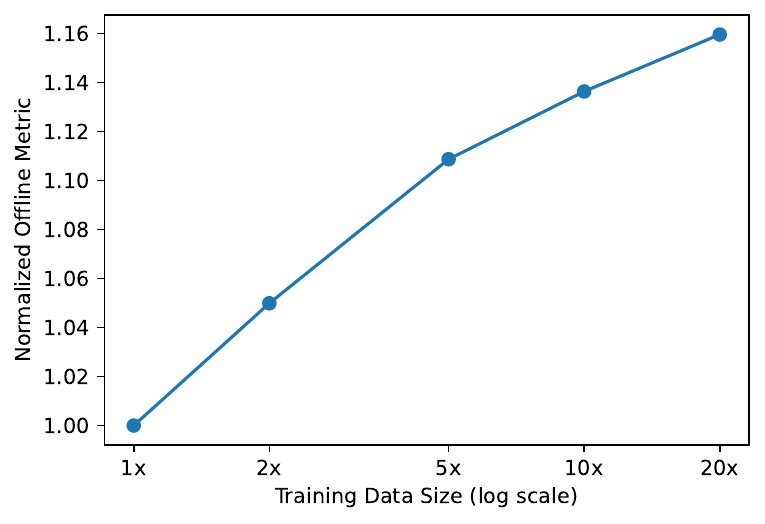}
  \caption{Phase‑2 data scaling for the $\sim$10B model.}
  \label{fig:data-scaling}
\end{figure}

\textbf{Data Scaling:} We first study how recommendation quality scales with the volume of Phase‑2 post‑training data. Holding other variables fixed, we train GenRec on datasets ranging from 1x to 20x the size of our smallest data size configuration. We run this study for two backbone sizes: a smaller model (on the order of $\sim$1B parameters) and a larger model (on the order of $\sim$10B parameters). For both model sizes, we observe that offline ranking metric MRR improves monotonically as we increase the amount of Phase‑2 data (Figure ~\ref{fig:data-scaling}). Although the absolute MRR of the $\sim$1B model is lower than that of the $\sim$10B model, it exhibits a similar scaling pattern to $\sim$10B model.

\textbf{Model Scaling:} We also examine how recommendation quality varies with backbone size. We post‑train a few GenRec variants with base models of different sizes (on the order of $\sim$1B to $\sim$10B parameters) under a fixed training budget (same GPU configuration and similar wall‑clock training time). Under this constraint, larger backbones consistently achieve higher offline MRR than smaller ones. 

\textbf{Quality-Cost Tradeoff in Phase 2:} While data and model scaling both improve quality, they also increase \textbf{training and serving costs}, especially for GenRec that is trained at a higher cadence and directly serves large‑scale online traffic.
We therefore explicitly search for Phase‑2 configurations that balance quality and cost: large enough to capture most of the incremental quality gains, but not so large that post‑training or serving becomes prohibitively expensive. In practice, we use the data‑ and model‑scaling curves, together with context‑length ablations, to identify a “sweet spot” on the quality–cost Pareto frontier that recovers most of the achievable quality improvement while staying within a given compute budget.

% \textbf{Data Efficiency Relative to the Production Ranker:} Although Phase‑2 quality improves as we add more post‑training data, GenRec still requires substantially less labeled interaction data than our traditional production ranker to reach comparable quality. In particular, when starting from a strong Netflix‑adapted foundational LLM (Phase 1), we find that Phase‑2 GenRec can match the production model’s offline and online metrics using 10-40x fewer Phase‑2 training examples, depending on the configuration. This marginal data efficiency shows that Phase 2 benefits from the user and content understanding already learned in Phase 1, and therefore does not need to “re‑learn” that knowledge from scratch. Because Phase 1 is updated at a relatively low cadence, achieving better data efficiency in the more frequently retrained Phase‑2 model is especially meaningful in practice.

\subsection{Impact of Phase 1 \& Phase 2 on Recommendation Task}

We quantify the contributions of Phase‑1 foundation training and Phase‑2 post‑training on offline ranking metrics.

\textbf{Phase 1. }Using the Phase‑1 foundational LLM as the base model improves offline ranking metrics by the order of \textbf{10-20\% } compared with using an off‑the‑shelf LLM as the backbone. This underscores the importance of the personalization and content‑understanding capabilities inherited from the Phase 1 foundational model. 

\textbf{Phase 2. }Phase‑2 post‑training provides a further roughly \textbf{35–50\%} gain in offline ranking metrics when evaluated on the Phase-1 training cutoff date (i.e. when Phase‑1 model is the freshest). This highlights the value of task‑specific adaptation, including domain‑specific data, rewards, and verbalizations. In addition, the relative benefit of Phase 2 grows over time: the improvement rises to roughly 80\% after two weeks, which is expected. This growing gain reflects both the strength of task‑specific adaptation and the increasing staleness of the less frequently updated Phase‑1 backbone, for example with respect to shifting content popularity trends and evolving member interests.

\begin{table}
  \caption{Impact of Phase 1 vs. Phase 2 training}
  \label{tab:phase-impact}
  \begin{tabular}{lp{0.66\columnwidth}}   
    \toprule
     & Offline ranking metrics (MRR) \\
    \midrule
    Impact of Phase 1 & \textbf{+ 10-20\%} compared to using an OSS model as base \\
    Impact of Phase 2 & \textbf{+ 35-50\%} compared to a freshly trained Phase 1, with the gain increasing over time \\
    \bottomrule
  \end{tabular}
\end{table}

\subsection{Context Length Optimization}

The context window in an LLM is both a quality driver and a cost driver: longer verbalizations expose more information about user behavior and context but increase training and serving cost. We therefore study how context length and verbosity affect recommendation performance and explicitly search for compact verbalizations that preserve quality. We build on the “Input Verbalization and Context Engineering” heuristics in a few steps. 

\begin{figure}[h!]
  \centering
  \includegraphics[width=\linewidth]{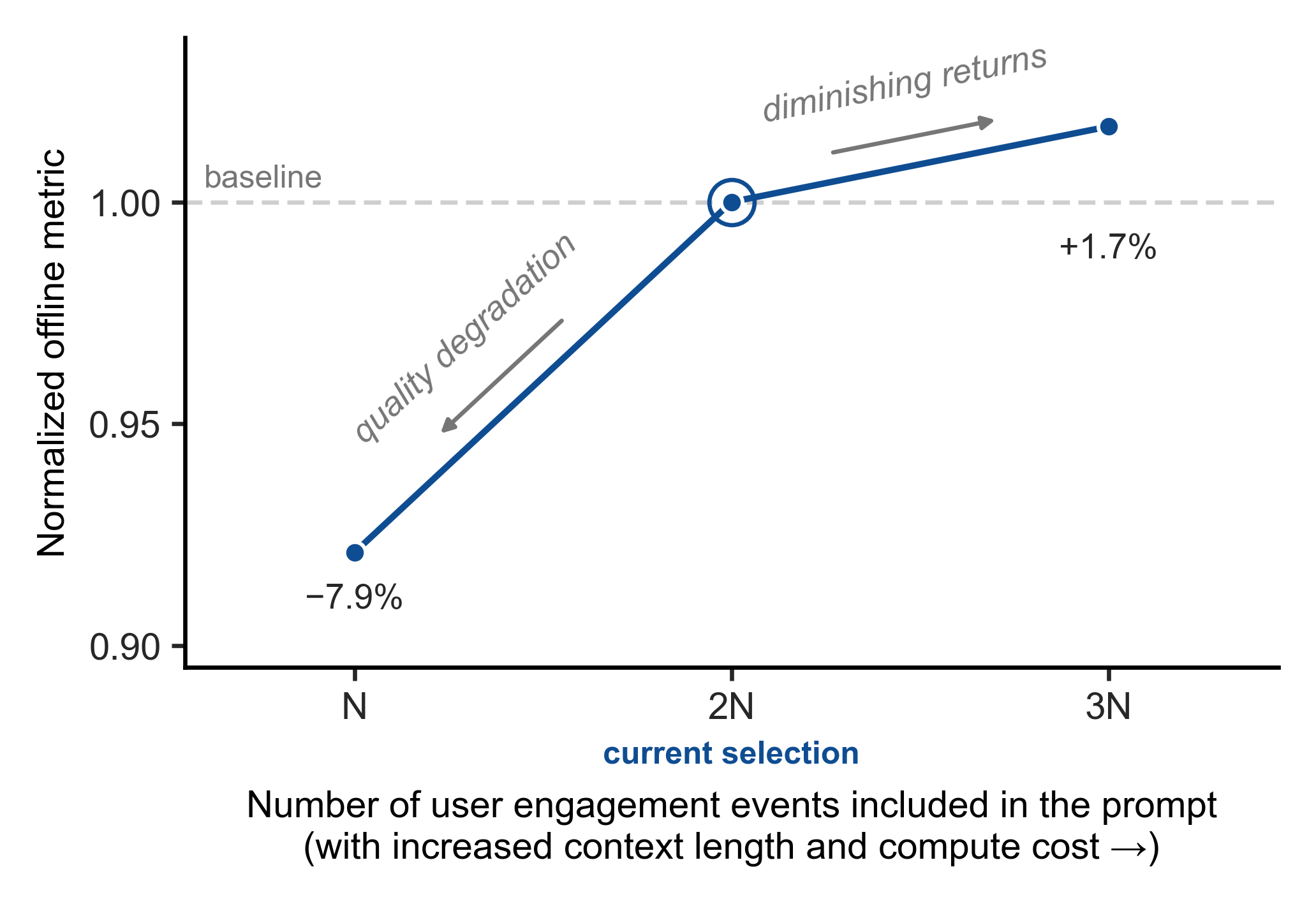}
  \caption{Normalized offline ranking metric (MRR) vs. number of user engagement events included in the prompt. The dashed line marks the baseline elbow point: increasing the number of events beyond this yields diminishing returns.}
  \Description{Chart illustrating the cost-quality tradeoff as verbalization token budget varies, showing that a compact representation preserves quality while reducing inference cost.}
  \label{fig:context-engineering}
\end{figure}

\begin{itemize}
    \item \textbf{First, we perform event selection and compression.} We omit low‑signal or noisy engagements, compress repetitive activity, and retain high‑signal interactions. This yields a cleaned, ordered sequence of engagements that are candidates for inclusion in the context.
    \item \textbf{Second, we search for an optimal history length to elaborate.} Starting from this cleaned sequence, we vary the number of historical events we keep in the prompt. Concretely, we sweep the number of retained engagements, and plot MRR against the number of engagement events. This reveals an elbow point: quality improves as we increase context length up to that point, after which additional tokens yield only marginal gains (Figure ~\ref{fig:context-engineering}).
    \item \textbf{Finally, we construct a concise and efficient verbalization.} For each retained event, we construct prompts with different levels of detail (e.g. different engagement details, simplifying wording, and removing few‑shot examples) and measure offline performance.
\end{itemize}

In our experiments, we observe the context length can be reduced to roughly \textbf{a third} of the original token budget (for example, from about 5,000 tokens down to about 1,700 tokens) with only negligible degradation in offline ranking metrics. This corresponds to a substantial reduction in context size—and therefore inference cost—while maintaining essentially the same model quality. Since GenRec is largely compute-bound with serving cost approximately proportional to the context length,  we observed the serving cost is reduced to roughly \textbf{a third} of original cost as well.

%These results show that we can identify a sweet spot for context length: a compact representation that is sufficient to match the performance of a much longer, noisier history. This validates verbalization compaction as a robust mechanism for balancing the quality–cost trade‑off and underscores that context engineering is critical to efficient performance in LLM‑backed recommenders.

\section{Discussion: LLM-Native Recommendation}

GenRec is an initial step toward a more LLM‑centric recommendation stack at Netflix. Moving from traditional recommendation systems to LLM‑based ones is not just “swapping the ranker with a transformer”: it changes how we represent users and content, how we structure models and training, and how we design serving infrastructure. Below, we outline several of these changes and how they manifest in GenRec. Taken together, they move our recommendation stack closer to the broader LLM paradigm.

\subsection{From Feature Engineering to Context Engineering}

 Traditional systems are built around large number of handcrafted features, supported by substantial infrastructure for aggregation, freshness, interaction design, and leakage control. In contrast, LLM‑centric systems prioritize constructing rich textual contexts from raw interaction sequences, content metadata, context, tool outputs, or user memory. The “prompt” becomes the new feature vector. Modeling effort shifts from designing individual features to context engineering: deciding which signals to include, how to summarize them, how far back in time to go, and how to encode them within a finite token budget. Our experiments on verbalization compaction illustrate this shift: careful context design can preserve quality while dramatically reducing serving cost.

\subsection{From Customized Architectures to Foundation Backbones}

 Historically, recommender systems have relied on a wide variety of bespoke architectures—two‑tower models, DLRM‑style feature interaction networks, custom attention blocks, and large multi‑task setups. LLM‑centric systems instead standardize on transformer backbones inherited from a foundation model. Innovation moves up a level: from per‑task architecture design to questions of data, scaling laws, post‑training strategy, and inference optimization. In GenRec, we leverage the same LLM backbone as the foundational LLM, rather than designing a new architecture from scratch. The LLM backbone also enables flexible recommendation steering through natural‑language inputs to support new member experiences.

\subsection{Scaling Laws as a Design Guide}

Classical RecSys stacks can hit diminishing returns due to sparse IDs, heavily engineered objectives, and task‑specific architectures. In an LLM‑backed setting, the recommendation model shares its backbone with a pre‑trained LLM and inherits its data and model scaling behavior. Our studies show clear scaling trends in both data and model size: performance improves monotonically as we add more Phase‑2 training data, and larger models consistently outperform smaller ones under fixed training budgets. This brings RecSys closer to the broader LLM paradigm, where scaling laws are treated as a central design guide rather than an after‑the‑fact observation.

\subsection{Pre-Training, Post-Training, and Reward Alignment}

Instead of building each recommender from scratch, an LLM‑centric approach starts from a pre‑trained foundation model and then uses task‑specific post‑training and reward alignment to adapt this shared backbone. In GenRec, Phase 1 provides a Netflix‑aware foundation LLM, and Phase 2 performs lightweight, recommendation‑specific adaptation with labels, rewards, and verbalizations. This pattern makes it easier to scale innovation across applications: multiple use cases can reuse the same foundation and infrastructure, rather than re‑implementing similar capabilities in separate, siloed systems.

\subsection{From RecSys Infrastructure to LLM Infrastructure}

LLM‑backed recommenders must process longer user‑history contexts, which makes serving efficiency a central design concern. At the same time, LLMs are inherently expensive to train and serve, so cost optimization must shape choices around model size, context length, and decoding strategy. Techniques such as KV‑caching, prefix caching, and prefill‑only inference become key levers for keeping inference viable within cost budgets. As a result, the online serving stack for recommendation increasingly resembles modern LLM infrastructure—GPU‑accelerated, vLLM/Triton‑based, with careful batching and caching—rather than classic RecSys infrastructure built around MLPs or factorization models.

\section{Conclusions \& Future Work}

We presented GenRec, an LLM‑backed recommendation ranker at Netflix, and described how we adapt an in‑house foundational LLM into a ranker suitable for production traffic on batch-compute surfaces. By verbalizing user histories and context, adding a catalog‑aware ranking head, and training with multiple reward signals, we obtain an LLM-backed recommender that, despite using far less Phase2-2 labels and explicit input signals than our long-standing production ranker, achieves statistically significant improvement over it on both short-term and long-term online metrics in the settings we study.

Overall, GenRec represents an initial step toward a more LLM‑centric recommendation stack at Netflix. Our results suggest that this is a viable and promising path for large‑scale personalization in at least some scenarios, provided we continue to manage cost and infrastructure complexity carefully and to evaluate quality trade‑offs rigorously against strong non‑LLM baselines.

\section{Acknowledgments}

GenRec is the result of close collaboration among multiple teams and organizations across Netflix. The contributors to this work (in alphabetical order):

\begin{itemize}
    \item \textbf{AI for members: }Arjun Rao, Ashish Rastogi, Baolin Li, Fernando Amat Gil, Grace Huang, Justin Basilico, Kamelia Aryafar, Linas Baltrunas, Moumita Bhattacharya, Ogheneovo Dibie, Rein Houthooft, Shradha Sehgal, Sejoon Oh, Sergi Perez, Sourabh Medapati, Thea Wang, Yaochen Zhu, Yesu Feng, Ying Li, Yun Li, Yucheng Shi, Yunan Hu
    \item \textbf{AI platform and serving: }Abhishek Agrawal, Adam Singer, Binh Tang, Daneo Zhang, Derek Olejnik, Ed Maddox, Erik Osheim, Lingyi Liu, Liping Peng, Meghana Chilukuri, Nicolas Hortiguera, Shaojing Li, ZQ Zhang
    \item \textbf{Product: }Ilke Kaya, Michelle Kislak, Scarlet Chen, Si Cheng
\end{itemize}

\bibliographystyle{ACM-Reference-Format}
\bibliography{genrecs}

\end{document}